\documentclass{article}
\usepackage{amsfonts}
\usepackage{amsmath}
\usepackage{cite}

\input{tcilatex}

\begin{document}

\title{ Auto-correlation Function Analysis of Scattered Light Intensity at
Different Scattering Angles}
\author{Yong Sun{\thanks{%
Email: ysun2002h@yahoo.com.cn}}}
\maketitle

\begin{abstract}
In this work, the effects of the scattering angle on the nonexponentiality
of the normalized time auto-correlation function of the scattered light
intensity $g^{\left( 2\right) }\left( \tau \right) $ are investigated using
dilute Poly($N$-isopropylacrylamide) microgel and standard polystyrene latex
samples in dispersion respectively. The results show that the influences of
the scattering angle on the deviation between an exponentiality and $%
g^{\left( 2\right) }\left( \tau \right) $ are small. With the assistance of
the simulated data of $g^{\left( 2\right) }\left( \tau \right) $, the
effects of the particle size distribution and scattering angle on the
deviation between an exponentiality and $g^{\left( 2\right) }\left( \tau
\right) $ are explored. The analysis reveals that the nonexponentiality of $%
g^{\left( 2\right) }\left( \tau \right) $ is determined by the particle size
distribution and scattering angle. In general, the influences of the
particle size distribution are small on the nonexponentiality of $g^{\left(
2\right) }\left( \tau \right) $ and very large on the initial slope of the
logarithm of $g^{\left( 2\right) }\left( \tau \right) $ and the effects of
the scattering angle are determined by the particle size distribution and
mean particle size. Under some conditions, the deviation between an
exponentiality and $g^{\left( 2\right) }\left( \tau \right) $ is greatly
influenced by the scattering angle. The values of the apparent hydrodynamic
radius are also determined by the particle size distribution and scattering
angle. The apparent hydrodynamic radius and its distribution obtained using
the cumulants method are different from the hydrodynamic radius and its
distribution.
\end{abstract}

\section{INTRODUCTION}

Light scattering is a widely used technique to measure the properties of
particles in colloidal dispersion systems. One of the main applications of
the dynamic light scattering (DLS) technique is to measure the sizes of
spherical particles in liquid suspension. The standard method of cumulants%
\cite{re1,re2,re3,re4} has been used to measure the hydrodynamic radius, or
more strictly apparent hydrodynamic radius $R_{h,app}$ \cite{re5} of
particles from the normalized time auto-correlation function of the
scattered light intensity $g^{\left( 2\right) }\left( \tau \right) $ with
the assistance of the Einstein-Stokes relation, where $\tau $ is the delay
time. In order to obtain the effective diffusion coefficient\cite{re6} or
the apparent hydrodynamic radius\cite{re5} to detect small poly-dispersities
for large particles, DLS technique is endeavored to use at different
scattering angles.

In the previous work\cite{re7,re8,re9}, the particle size information
included in the static and dynamic light scattering data and the influences
of particle size distribution on the deviation between an exponentiality and 
$g^{\left( 2\right) }\left( \tau \right) $ at a scattering angle of 30$^{%
\text{o}}$ have been explored. The particle size distributions obtained
using different techniques have been compared. In this work, the effects of
the scattering angle on the nonexponentiality of $g^{\left( 2\right) }\left(
\tau \right) $ are investigated using dilute Poly($N$-isopropylacrylamide)
microgel and standard polystyrene latex samples in dispersion respectively.
The results show that the effects of the scattering angle on the deviation
between an exponentiality and $g^{\left( 2\right) }\left( \tau \right) $ are
small. Since the method of cumulants measures the particle size distribution
from the nonexponentiality of $g^{\left( 2\right) }\left( \tau \right) $ at
a single scattering angle, the simulated data were thus used to explore the
effects of the particle size distribution on the difference between them at
different scattering angles. The analysis reveals that the deviation between
an exponentiality and $g^{\left( 2\right) }\left( \tau \right) $ is
determined by the particle size distribution and scattering angle. In
general, the influences of the particle size distribution are small on the
nonexponentiality of $g^{\left( 2\right) }\left( \tau \right) $ and very
large on the initial slope of the logarithm of $g^{\left( 2\right) }\left(
\tau \right) $ and the effects of the scattering angle are determined by the
particle size distribution and mean particle size. Under some conditions,
the deviations between an exponentiality and $g^{\left( 2\right) }\left(
\tau \right) $ are greatly influenced by the scattering angle. The values of
the apparent hydrodynamic radius are also determined by the particle size
distribution and scattering angle. The apparent hydrodynamic radius and its
distribution obtained using the cumulants method are different from the
hydrodynamic radius and its distribution.

\section{THEORY}

When the Rayleigh-Gans-Debye (RGD) approximation is valid, the normalized
time auto-correlation function of the electric field of the scattered light $%
g^{\left( 1\right) }\left( \tau \right) $ for dilute poly-disperse
homogeneous spherical particles in dispersion is given by

\begin{equation}
g^{\left( 1\right) }\left( \tau \right) =\frac{\int_{0}^{\infty
}R_{s}^{6}P\left( q,R_{s}\right) G\left( R_{s}\right) \exp \left(
-q^{2}D\tau \right) dR_{s}}{\int_{0}^{\infty }R_{s}^{6}P\left(
q,R_{s}\right) G\left( R_{s}\right) dR_{s}},  \label{Grhrs}
\end{equation}
where $R_{s}$ is the static radius, $D$ is the diffusion coefficient, $\ q=%
\frac{4\pi }{\lambda }n_{s}\sin \frac{\theta }{2}$ is the scattering vector, 
$\lambda $\ is the wavelength of the incident light in vacuo, $n_{s}$ is the
solvent refractive index, $\theta $ is the scattering angle, $G\left(
R_{s}\right) $ is the number distribution of particle sizes and the form
factor $P\left( q,R_{s}\right) $ is

\begin{equation}
P\left( q,R_{s}\right) =\frac{9}{q^{6}R_{s}^{6}}\left( \sin \left(
qR_{s}\right) -qR_{s}\cos \left( qR_{s}\right) \right) ^{2}.  \label{factor}
\end{equation}

\noindent From the Einstein-Stokes relation

\begin{equation}
D=\frac{k_{B}T}{6\pi \eta _{0}R_{h}},
\end{equation}
where $\eta _{0}$, $k_{B}$ and $T$ are the viscosity of the solvent,
Boltzmann's constant and absolute temperature respectively, the hydrodynamic
radius $R_{h}$ can be obtained.

Traditionally the cumulants is a standard method to measure the particle
size distribution from the DLS data $g^{\left( 2\right) }\left( \tau \right)$%
. In this work, the following equation was used to analyze the DLS data to
the second moment

\begin{equation}
g^{\left( 2\right) }\left( \tau \right) =1+\beta \exp \left( -2\left\langle
\Gamma \right\rangle \tau \right) \left( 1+\mu _{2}\tau ^{2}\right)
\label{cum}
\end{equation}
where $\left\langle \Gamma \right\rangle=q^2D_e\left(q\right)$ is the
average decay rate, $D_e\left(q\right)$ is the effective diffusion
coefficient, $\mu _2$ is the second moment and $\beta$ is a constant that
depends on the experimental geometry for a given experimental measurement.
The apparent hydrodynamic radius $R_{h,app}$ can be obtained from $%
D_e\left(q\right)$

\begin{equation}
R_{h,app}=\frac{k_{B}T}{6\pi \eta _{0}D_e} .  \label{Rhapp}
\end{equation}

\noindent The relative width of the apparent hydrodynamic radius
distribution is\cite{re10} 
\begin{equation}
\frac{Width}{R_{h,app}}=\frac{\sqrt{\mu _2}}{\left\langle \Gamma
\right\rangle}.  \label{Rhwidth}
\end{equation}

\noindent If the first cumulant is used, the value of the apparent
hydrodynamic radius $R_{h,app}$ at given scattering angle and delay time $%
\tau$ can be calculated directly using the static particle size information
and the relationship between the static and hydrodynamic radii. If the DLS
data during the delay time range $\tau _1$ and $\tau _2$ are chosen to
obtain $R_{h,app}$ at a given scattering angle, the average value of
apparent hydrodynamic radius can be calculated using the following equation

\begin{equation}
R_{h,app}\left( e^{-\frac{q^{2}k_{B}T\tau _{1}}{6\pi \eta _{0}R_{h,app}}%
}-e^{-\frac{q^{2}k_{B}T\tau _{2}}{6\pi \eta _{0}R_{h,app}}}\right) =\frac{%
\int_{0}^{\infty }R_{h}R_{s}^{6}P\left( q,R_{s}\right) G\left( R_{s}\right)
\left( e^{-\frac{q^{2}k_{B}T\tau _{1}}{6\pi \eta _{0}R_{h}}}-e^{-\frac{%
q^{2}k_{B}T\tau _{2}}{6\pi \eta _{0}R_{h}}}\right) dR_{s}}{\int_{0}^{\infty
}R_{s}^{6}P\left( q,R_{s}\right) G\left( R_{s}\right) R_{s}}.
\label{Rhallcal}
\end{equation}

\section{EXPERIMENT}

The SLS and DLS data were measured using the instrument built by ALV-Laser
Vertriebsgesellschaft m.b.H (Langen, Germany). It utilizes an ALV-5000
Multiple Tau Digital Correlator and a JDS Uniphase 1145P He-Ne laser to
provide a 23 mW vertically polarized laser at wavelength of 632.8 nm.

In this work, two kinds of samples were used. One is PNIPAM submicron
spheres and the other is standard polystyrene latex spheres. The samples
used in this work have been detailed before\cite{re7}. The four PNIPAM
microgel samples PNIPAM-0, PNIPAM-1, PNIPAM-2 and PNIPAM-5 were named
according to the molar ratios $n_{B}/n_{N}$ of cross-linker $N,N^{\prime }$%
-methylenebisacrylamide over $N$-isopropylacrylamide. The sulfate
polystyrene latex with a normalized mean radius of 33.5 nm and
surfactant-free sulfate polystyrene latex of 55 nm were named Latex-1 and
Latex-2 respectively.

\section{DATA ANALYSIS}

In this section, the nonexponentiality of $g^{\left( 2\right) }\left( \tau
\right) $ is investigated using the standard polystyrene latex and PNIPAM
samples at different scattering angles.

\subsection{Small particles}

For small particles, the standard DLS technique can be used at any given
scattering angle. For narrow particle size distributions, the effects of the
form factor can be neglected, the particle size information obtained at
different scattering angles should be the same. The particle sizes and
distributions of the standard polystyrene latex samples Latex-1 and Latex-2
provided by the supplier are small and narrow respectively. The deviation
between an exponentiality and $g^{\left( 2\right) }\left( \tau \right) $ are
investigated at a temperature of 298.5 K and scattering angles 30$^{\text{o}%
} $, 40$^{\text{o}}$, 50$^{\text{o}}$, 60$^{\text{o}}$, 70$^{\text{o}}$, 80$%
^{\text{o}}$, 90$^{\text{o}}$, 100$^{\text{o}}$, 120$^{\text{o}}$, 130$^{%
\text{o}}$, 140$^{\text{o}}$ and 150$^{\text{o}}$, respectively. The
logarithm of the DLS data $\ln \left( g^{\left( 2\right) }\left( \tau
\right) -1\right) $ was plotted as a function of the delay time $\tau $,
respectively. The results for Latex-1 and Latex-2 measured at scattering
angles 30$^{\text{o}}$, 60$^{\text{o}}$, 90$^{\text{o}}$, 120$^{\text{o}}$
and 150$^{\text{o}}$ are shown in Figs. 1a and 1b respectively. Figure 1
shows clearly the nonexponentiality of $g^{\left( 2\right) }\left( \tau
\right) $ at a given scattering angle investigated is small. The values of
apparent hydrodynamic radius obtained at different scattering angles almost
are equal. For Latex-1 and Latex-2, the values are about 37.27 nm and 64.5
nm respectively.

For PNIPAM samples at high temperatures, the particle sizes and
distributions obtained using the SLS technique are small and narrow. The
nonexponentiality of $g^{\left( 2\right) }\left( \tau \right) $ for PNIPAM
samples at hight temperatures are investigated as the standard polystyrene
latex samples Latex-1 and Latex-2 at a temperature of 298.5 K. The results
for PNIPAM-1 and PNIPAM-2 at about a temperature of 312.6 K and scattering
angles 30$^{\text{o}}$, 60$^{\text{o}}$, 90$^{\text{o}}$, 120$^{\text{o}}$
and 150$^{\text{o}}$ are shown in Figs. 2a and 2b respectively. Figure 2
shows clearly the nonexponentiality of $g^{\left( 2\right) }\left( \tau
\right) $ at a given scattering angle investigated is small. The values of
apparent hydrodynamic radius obtained at different scattering angles almost
are equal. For PNIPAM-1 and PAIPAM-2, the values are about 137.8 nm and
136.5 nm respectively.

\subsection{Large particles}

For large particles, since most of scattering light is cancelled due to the
light interference in the vicinity of the scattered intensity minimum and
the scattered intensity at large scattering angles is influenced by the
reflected light, the information included in DLS data is complex. In order
to avoid the considerations for the situations, the nonexponentiality of $%
g^{\left( 2\right) }\left( \tau \right) $ is investigated at small
scattering angles before the emergence of the scattered intensity minimum.
For the samples PNIPAM-1 and PANIPAM-2 at a temperature of 302.2 K, the
scattered intensity minimums emerge at about the scattering angles 85$^{%
\text{o}}$ and 97$^{\text{o}}$ respectively, the deviations between an
exponentiality and $g^{\left( 2\right) }\left( \tau \right) $ are explored
before the scattering angle 75$^{\text{o}}$. The logarithm of the DLS data $%
\ln \left( g^{\left( 2\right) }\left( \tau \right) -1\right) $ was plotted
as a function of the delay time $\tau $. The results for PNIPAM-1 and
PNIPAM-2 at scattering angles 30$^{\text{o}}$, 50$^{\text{o}}$ and 70$^{%
\text{o}}$ are shown in Figs. 3a and 3b respectively. Figure 3 shows clearly
the nonexponentiality of $g^{\left( 2\right) }\left( \tau \right) $ at a
given scattering angle investigated is small. The values of apparent
hydrodynamic radius are a function of the scattering angle. For PNIPAM-1 the
values are about 320 nm to 310 nm and for PNIPAM-2 the values are from 300
nm to 295 nm during scattering angle range from 30$^{\text{o}}$ to 70$^{%
\text{o}}$.

\section{RESULTS AND DISCUSSION}

Because the distributions of the PNIPAM submicron and standard polystyrene
latex samples are narrow and the expected values of the DLS data calculated
based on the commercial and static particle size information are consistent
with the experimental data\cite{re7,re8}, the DLS simulated data were used
in order to explore the effects of particle size distribution on the
deviation between an exponentiality and $g^{\left( 2\right) }\left( \tau
\right) $ and the initial slope of the logarithm of $g^{\left( 2\right)
}\left( \tau \right) $ at different scattering angles in detail. The method
that produces the DLS simulated data has been detailed before\cite{re7}. In
this work, the number distribution of particle sizes is still chosen as a
Gaussian distribution

\begin{equation}
G\left( R_{s};\left\langle R_{s}\right\rangle ,\sigma \right) =\frac{1}{
\sigma \sqrt{2\pi }}\exp \left( -\frac{1}{2}\left( \frac{R_{s}-\left\langle
R_{s}\right\rangle }{\sigma }\right) ^{2}\right) ,
\end{equation}
where $\left\langle R_{s}\right\rangle $ is the mean static radius and $%
\sigma $ is the standard deviation related to the mean static radius.

The simulated data were produced using the information: the mean static
radius $\left\langle R_{s}\right\rangle $, standard deviation $\sigma $,
temperature $T$, viscosity of the solvent $\eta _{0}$, scattering angle $%
\theta $, wavelength of laser light $\lambda $, refractive index of the
water $n_{s}$ and constant $a=R_{h}/R_{s}$ were set to 50 nm, 10 nm,
300.49K, 0.8479 mPa$\cdot $S, 90$^{\text{o}}$, 632.8 nm, 1.332 and 1.1,
respectively. When the data of $\left( g^{\left( 2\right) }\left( \tau
\right) -1\right) /\beta $ were obtained, the 1\% statistical noises were
added and the random errors were set 3\%. Five simulated data were produced
respectively. The fit results for one of the DLS simulated data at different
delay time ranges using Eq. \ref{cum} with $\mu _{2}=0$ and $\mu _{2}\neq 0$
respectively are listed in Table \ref{data501090}.

\begin{table}[tbp]
\begin{center}
\begin{tabular}{|c|c|c|c|c|c|}
\hline
Delay time (s) & $\left\langle\Gamma\right\rangle _{first}\left(
s^{-1}\right) $ & $\chi ^{2}$ & $\left\langle\Gamma\right\rangle
_{two}\left( s^{-1}\right) $ & $\mu _{2}\left( s^{-2}\right) $ & $\chi ^{2}$
\\ \hline
2*10$^{-7}$ to 0.00151 & 1395.0$\pm $0.8 & 1.42 & 1420$\pm $3 & 46900$\pm $
4900 & 0.42 \\ \hline
2*10$^{-7}$ to 0.00166 & 1390.2$\pm $0.7 & 2.79 & 1421$\pm $2 & 50500$\pm $%
3400 & 0.41 \\ \hline
2*10$^{-7}$ to 0.00181 & 1389.6$\pm $0.6 & 2.86 & 1421$\pm $2 & 50300$\pm $%
3300 & 0.40 \\ \hline
2*10$^{-7}$ to 0.00196 & 1386.4$\pm $0.6 & 3.92 & 1421$\pm $2 & 50100$\pm $%
2700 & 0.39 \\ \hline
2*10$^{-7}$ to 0.00211 & 1384.7$\pm $0.5 & 4.51 & 1421$\pm $2 & 50200$\pm $%
2500 & 0.38 \\ \hline
2*10$^{-7}$ to 0.00226 & 1374.1$\pm $0.4 & 9.63 & 1423$\pm $2 & 53300$\pm $%
1900 & 0.39 \\ \hline
2*10$^{-7}$ to 0.00241 & 1373.3$\pm $0.3 & 9.72 & 1422$\pm $2 & 52400$\pm $%
1800 & 0.41 \\ \hline
\end{tabular}
\makeatletter
\par
\makeatother
\end{center}
\caption{The fit results of simulated data produced based on the mean static
radius 50 nm, standard deviation 10 nm and a scattering angle of 90$^\mathrm
o$ at different delay time ranges using Eq. \protect\ref{cum} with $\protect%
\mu_2=0$ and $\protect\mu_2 \neq 0$ respectively.}
\label{data501090}
\end{table}

The fit results of $\left\langle \Gamma \right\rangle $ and $\mu _{2}$ are
influenced by the delay time range being fit as shown in Table \ref%
{data501090}. When Eq. \ref{cum} was used to fit the simulated data produced
based on the mean static radius 50 nm and standard deviation 10 nm at a
scattering angle of 90$^{\text{o}}$ under the conditions of $\mu _{2}=0$ and 
$\mu _{2}\neq 0$ respectively, it was found that the uncertainties in
parameters decrease and $\left\langle \Gamma \right\rangle $ and $\mu _{2}$
stabilize as the delay time range is increased. From Eq. \ref{Rhwidth}, the
relative width of the apparent hydrodynamic radius distribution is about
0.16. This value is equal to that of the relative width of the apparent
hydrodynamic radius distribution obtained at a scattering angle of 30$^{%
\text{o}}$\cite{re9} and different from the relative width of this simulated
data 0.2. Fit results obtained using both procedures at the delay time range
2$\times 10^{-7}$ to 0.00196 s are shown in Fig. 4. For both the fit
results, the residuals $\left( y_{i}-y_{fit}\right) /\sigma _{i}$ are random
as the delay time is changed, where $y_{i}$, $y_{fit}$ and $\sigma _{i}$ are
the data, the fit value and the uncertainty in the data at a given delay
time $\tau _{i}$, respectively.

The fit results for other particle size distributions also have been
analyzed. The simulated data were produced using the same temperature $T$,
viscosity of the solvent $\eta _{0}$, wavelength of laser light $\lambda $,
mean static radius $\left\langle R_{s}\right\rangle $, constant $a$ and
refractive index of the water $n_{s}$. The standard deviations and
scattering angles $\theta $ were set to 3 nm, 5 nm, 15 nm, 20 nm and 25 nm
and 60$^{\text{o}}$, 90$^{\text{o}}$, 120$^{\text{o}}$ and 150$^{\text{o}}$,
respectively. Since the standard method of cumulants obtains the
distribution of apparent hydrodynamic radius from the deviation between an
exponentiality and $g^{\left( 2\right) }\left( \tau \right) $ at a single
scattering angle, the simulated data were used to investigated the effects
of the particle size distributions on the nonexponentiality of $g^{\left(
2\right) }\left( \tau \right) $ at different scattering angles. The
logarithm of the simulated data produced without noises and errors was
plotted as a function of the delay time $\tau $. All results for the
standard deviations 3 nm, 10 nm, 20 nm, 25 nm and mean static radius 50 nm
are shown in Fig. 5a for a scattering angle of 90$^{\text{o}}$ and in Fig.
5b for a scattering angle of 150$^{\text{o}}$. Figure 5 shows the effects of
the standard deviation are small on the nonexponentiality of $g^{\left(
2\right) }\left( \tau \right) $ and large on the initial slope of the
logarithm of $g^{\left( 2\right) }\left( \tau \right) $ $\left\langle \Gamma
\right\rangle $ at scattering angles 90$^{\text{o}}$ and 150$^{\text{o}}$.

Since the particle size is an important quantity obtained using the DLS
technique, the effects of the particle size distribution on the apparent
hydrodynamic radius were thus investigated. The values of the apparent
hydrodynamic radius obtained using Eqs. \ref{cum} and \ref{Rhapp} with $\mu
_{2}=0$ and $\mu _{2}\neq 0$, and Eq. \ref{Rhallcal} respectively for the
simulated data produced using the mean static radius 50 nm and different
standard deviations at scattering angles 90$^{\text{o}}$ and 150$^{\text{o}}$
are shown in Table \ref{distribution}. From the relationship $%
a=R_{h}/R_{s}=1.1$, the mean hydrodynamic radius $\left\langle
R_{h}\right\rangle $ is 55 nm.

\begin{table}[tbp]
\begin{center}
\begin{tabular}{|c|c|c|c|c|c|c|}
\hline
& \multicolumn{3}{|c|}{Scattering angle 90$^{\text{o}}$} & 
\multicolumn{3}{|c|}{Scattering angle 150$^{\text{o}}$} \\ \hline
$\sigma /\left\langle R_{s}\right\rangle $ & $R_{h,app1}$(nm) & $R_{h,app2}$%
(nm) & $R_{cal}$(nm) & $R_{h,app1}$(nm) & $R_{h,app2}$(nm) & $R_{cal}$(nm)
\\ \hline
0.06 & 56.1$\pm $0.2 & 56.1$\pm $0.1 & 56.0 & 56.0$\pm $0.1 & 55.8$\pm $0.6
& 56.0 \\ \hline
0.1 & 57.8$\pm $0.2 & 57.2$\pm $0.6 & 57.7 & 57.8$\pm $0.2 & 57.6$\pm $0.6 & 
57.5 \\ \hline
0.2 & 65.1$\pm $0.1 & 64.3$\pm $0.5 & 64.7 & 64.3$\pm $0.4 & 63.1$\pm $0.4 & 
64.0 \\ \hline
0.3 & 74.4$\pm $0.5 & 72.6$\pm $0.6 & 73.8 & 72.6$\pm $0.5 & 70.4$\pm $0.6 & 
72.2 \\ \hline
0.4 & 84.7$\pm $0.5 & 82.1$\pm $0.7 & 83.7 & 81.5$\pm $0.3 & 79$\pm $1. & 
80.5 \\ \hline
0.5 & 94.6$\pm $0.4 & 90.7$\pm $0.8 & 93.53 & 90.2$\pm $0.5 & 86.3$\pm $0.9
& 88.7 \\ \hline
\end{tabular}
\makeatletter
\par
\makeatother
\end{center}
\caption{Values of $R_{h,app}$ obtained using Eqs. \protect\ref{cum} and 
\protect\ref{Rhapp} with $\protect\mu_2=0$ and $\protect\mu_2 \neq 0$, and
Eq. \protect\ref{Rhallcal} for the simulated data produced using the mean
static radius 50 nm and different standard deviations at scattering angles 90%
$^\mathrm o$ and 150$^\mathrm o$, respectively.}
\label{distribution}
\end{table}

The results in Table \ref{distribution} show that the value of the apparent
hydrodynamic radius is greatly influenced by the particle size distribution.
The part of apparent hydrodynamic radius represents the effects of particle
size distributions. The wider the particle size distribution, the larger the
value of the apparent hydrodynamic radius. The consistency between the value
calculated from Eq. \ref{Rhallcal} and the result obtained using the first
cumulant also shows the deviations between the exponentiality and $g^{\left(
2\right) }\left( \tau \right) $ at scattering angles 90$^{\text{o}}$ and 150$%
^{\text{o}}$ respectively are small even for very wide distribution like the
relative width distribution 50\%. The difference between the results
obtained using the first and first two cumulants is influenced by the
particle size distribution. For narrow distributions, they are almost equal.
For a wide distribution like 50\%, the difference is less than 5\%. The
relative width of apparent hydrodynamic radius obtained from the deviation
between an exponentiality and $g^{\left( 2\right) }\left( \tau \right) $ is
about 24\% at a scattering angle of 90$^{\text{o}}$ and 27\% at a scattering
angle of 150$^{\text{o}}$ for the simulated data produced using the relative
width of hydrodynamic radius 50\%.

Comparing the results of apparent hydrodynamic radius obtained at scattering
angles 30$^{\text{o}}$\cite{re9}, 90$^{\text{o}}$ and 150$^{\text{o}}$, for
narrow particle size distributions the values of apparent hydrodynamic
radius almost do not depend on the scattering angle and for wide particle
size distributions the values are a function of the scattering angle.
However, for any situation of the simulated data produced using the mean
static radius 50 nm, the deviations between an exponentiality and $g^{\left(
2\right) }\left( \tau \right) $ are small. In order to explore the question
further, the new simulated data were produced using the same temperature $T$%
, viscosity of the solvent $\eta _{0}$, wavelength of laser light $\lambda $
and refractive index of the water $n_{s}$. The mean static radius $%
\left\langle R_{s}\right\rangle $ and constant $a$ were set to 120 nm and
1.2 respectively. The standard deviations and scattering angles were set to
8, 12, 24, 36, 48, 60 nm and 30$^{\text{o}}$, 60$^{\text{o}}$, 90$^{\text{o}%
} $, 120$^{\text{o}}$, 150$^{\text{o}}$, respectively.

The simulated data were thus used to explore the nonexponentiality of $%
g^{\left( 2\right) }\left( \tau \right) $ at different scattering angles.
The logarithm of the simulated data produced without noises and errors was
plotted as a function of the delay time $\tau $. All results at scattering
angles 30$^{\text{o}}$ and 90$^{\text{o}}$ for the standard deviations 8 nm,
24 nm, 48 nm, 60 nm and mean static radius 120 nm are shown in Figs. 6a and
6b respectively. Figure 6 shows the effects of the standard deviation are
small on the nonexponentiality of $g^{\left( 2\right) }\left( \tau \right) $
and large on the initial slope of the logarithm of $g^{\left( 2\right)
}\left( \tau \right) $ $\left\langle \Gamma \right\rangle $ at scattering
angles 30$^{\text{o}}$ and 90$^{\text{o}}$.

The effects of the particle size distribution on the apparent hydrodynamic
radius were thus explored at a scattering angle of 90$^{\text{o}}$. The
values of the apparent hydrodynamic radius obtained using Eqs. \ref{cum} and %
\ref{Rhapp} with $\mu _{2}=0$ and $\mu _{2}\neq 0$, and Eq. \ref{Rhallcal}
respectively for the simulated data produced using the mean static radius
120 nm and different standard deviations are shown in Table \ref%
{distribution120}. From the relationship $a=R_{h}/R_{s}=1.2$, the mean
hydrodynamic radius $\left\langle R_{h}\right\rangle $ is 144 nm.

\begin{table}[tbp]
\begin{center}
\begin{tabular}{|c|c|c|c|}
\hline
$\sigma /\left\langle R_{s}\right\rangle $ & $R_{h,app1}\left( nm\right) $ & 
$R_{h,app2}\left( nm\right) $ & $R_{cal}\left( nm\right) $ \\ \hline
0.07 & 146.3$\pm $0.2 & 146.7$\pm $0.7 & 145.9 \\ \hline
0.1 & 148.5$\pm $0.3 & 148$\pm $1 & 148.2 \\ \hline
0.2 & 158.9$\pm $0.8 & 156.9$\pm $0.8 & 157.7 \\ \hline
0.3 & 168.5$\pm $0.7 & 164.2$\pm $0.9 & 167.0 \\ \hline
0.4 & 175.7$\pm $0.9 & 170.$\pm $1. & 174.1 \\ \hline
0.5 & 184.0$\pm $0.7 & 177$\pm $2 & 181.7 \\ \hline
\end{tabular}
\makeatletter
\par
\makeatother
\end{center}
\caption{Values of $R_{h,app}$ obtained using Eqs. \protect\ref{cum} and 
\protect\ref{Rhapp} with $\protect\mu_2=0$ and $\protect\mu_2 \neq 0$, and
Eq. \protect\ref{Rhallcal} for the simulated data produced using the mean
static radius 120 nm and different standard deviations at a scattering angle
of 90$^\mathrm o$.}
\label{distribution120}
\end{table}

The results in Table \ref{distribution120} also show that the value of the
apparent hydrodynamic radius is greatly influenced by the particle size
distribution. The part of apparent hydrodynamic radius represents the
effects of particle size distributions. The wider the particle size
distribution, the larger the value of the apparent hydrodynamic radius. The
consistency between the value calculated from Eq. \ref{Rhallcal} and the
result obtained using the first cumulant also shows the deviations between
the exponentiality and $g^{\left( 2\right) }\left( \tau \right) $ at a
scattering angle of 90$^{\text{o}}$ are small even for very wide
distribution like the relative width distribution 40\%, respectively. The
relative width of apparent hydrodynamic radius obtained from the deviation
between an exponentiality and $g^{\left( 2\right) }\left( \tau \right) $ is
about 27\% for the simulated data produced using the relative width of
hydrodynamic radius 50\% at a scattering angle of 30$^{\text{o}}$ and 26\%
at a scattering angle of 90$^{\text{o}}$. Meanwhile, according to Eq. \ref%
{Grhrs}, the apparent hydrodynamic radius obtained using the cumulants
method is a function of the scattering angle. In order to investigate the
effects of the scattering angle on the apparent hydrodynamic radius, the
initial slopes of $g^{\left( 2\right) }\left( \tau \right) $ at different
scattering angles were divided by the square of the scattering vectors
respectively. The results for the simulated data produced based on the mean
static radius 50 nm with the standard deviations 3 nm, 10 nm and 25 nm and
for the simulated data produced based on the mean static radius 120 nm with
the standard deviations 8 nm, 24 nm and 60 nm at scattering angles 30$^{%
\text{o}}$, 90$^{\text{o}}$ and 150$^{\text{o}}$ are shows in Figs. 7a and
7b, respectively. Figure 7a shows the values of the apparent hydrodynamic
radius do not depend on the scattering angle for the simulated data produced
based on the mean static radius 50 nm with the standard deviations 3 nm and
10 nm, and are a function of the scattering angle for the simulated data
obtained using the mean static radius 50 nm and standard deviation 25 nm.
Figure 7b shows the values of the apparent hydrodynamic radius do not depend
on the scattering angle for the simulated data obtained using the mean
static radius 120 nm and standard deviation 8nm and are a function of the
scattering angle for the simulated data produced based on the mean static
radius 120 nm and standard deviation 24 nm. Figure 7b also reveals that the
value obtained from the simulated data produced using the mean static radius
120 nm and standard deviations 24 nm at a scattering angle of 150$^{\text{o}%
} $ is equal to that obtained from the simulated data produced based on the
mean static radius 120 nm and standard deviations 8 nm and the
nonexponentiality of the DLS data produced using the mean static radius 120
nm and standard deviation 60 nm is very large at a scattering angle of 150$^{%
\text{o}}$.

In order to investigate further the effects of the scattering angle on the
nonexponentiality of the DLS data produced using the static radius 120 nm
and standard deviation 60 nm at scattering angles 30$^{\text{o}}$, 60$^{%
\text{o}}$, 90$^{\text{o}}$, 120$^{\text{o}}$ and 150$^{\text{o}}$, the
plots of $\ln \left( \left( g^{\left( 2\right) }\left( \tau \right)
-1\right) /\beta \right) $ as a function of the delay time $\tau $ are shown
in Fig. 8.

Figure 8 reveals the effects of the scattering angle on the
nonexponentiality of the DLS data are large. At scattering angles 30$^{\text{%
o}}$ and 60$^{\text{o}}$, the plots of $\ln \left( \left( g^{\left( 2\right)
}\left( \tau \right) -1\right) /\beta \right) $ as a function of the delay
time $\tau $ are consistent with the lines respectively and at scattering
angles 90$^{\text{o}}$, 120$^{\text{o}}$ and 150$^{\text{o}}$, the plots of $%
\ln \left( \left( g^{\left( 2\right) }\left( \tau \right) -1\right) /\beta
\right) $ as a function of the delay time $\tau $ deviate clearly from the
lines respectively. According to Eq. \ref{Grhrs}, two terms $P\left(
q,R_{s}\right) $ and $\exp \left( -q^{2}D\tau \right) $ include the
parameter of the scattering angle. In order to investigate the effects of
the form factor on the deviation between an exponentiality and $g^{\left(
2\right) }\left( \tau \right) $ , $P\left( q,R_{s}\right) $ was set to 1.
The simulated data were produced again at scattering angles 30$^{\text{o}}$,
60$^{\text{o}}$, 90$^{\text{o}}$, 120$^{\text{o}}$ and 150$^{\text{o}}$. $%
\ln \left( \left( g^{\left( 2\right) }\left( \tau \right) -1\right) /\beta
\right) $ was plotted as a function of the delay time $\tau $, respectively.
The results are shown in Fig. 9. Comparing Fig. 8 with Fig. 9, the
nonexponentiality of the DLS data is influenced obviously by $P\left(
q,R_{s}\right) $.

Figure 7 also reveals that the results obtained using the cumulants method
from the DLS data cannot be determined even if the plots of $\ln \left(
\left( g^{\left( 2\right) }\left( \tau \right) -1\right) /\beta \right) $ as
a function of the delay time $\tau $ are consistent with lines,
respectively. For example, the values of apparent hydrodynamic radius are 219%
$\pm $2 nm and 244$\pm $2 nm at a scattering angle of 30$^{\text{o}}$ and
175.7$\pm $0.9 nm and 184.0$\pm $0.7 nm at a scattering angle of 90$^{\text{o%
}}$ for the simulated data produced based on the mean static radius 120 nm
with the standard deviations 48 and 60 nm respectively. In order to
eliminate the effects of the scattering angle, the value of apparent
hydrodynamic radius can be obtained by approximating the scattering angle 0.
The calculated values obtained using intensity-weighted average diffusion
coefficient\cite{re6} were used to study this question in order to discuss
simply. The apparent hydrodynamic radius is calculated using the following
equation

\begin{equation}
R_{h,app}=\frac{\int R_{s}^{6}P\left( q,R_{s}\right) G\left( R_{s}\right)
dR_{s}}{\int R_{s}^{6}P\left( q,R_{s}\right) G\left( R_{s}\right)
/R_{h}dR_{s}}.  \label{RhappP}
\end{equation}

The values calculated using Eq. \ref{RhappP} are listed in Table \ref%
{averageD} for the simulated data produced based on the mean static radius
50 nm and different standard deviations at scattering angles 0, 30$^{\text{o}%
}$, 60$^{\text{o}}$, 90$^{\text{o}}$ and 120$^{\text{o}}$, respectively. The
results show the values of apparent hydrodynamic radius are a function of
the scattering angle and particle size distribution. Although the effects of
the scattering angle are eliminated at a scattering angle of 0, the values
of apparent hydrodynamic radius still are determined by the particle size
distribution. Because the distribution of apparent hydrodynamic radius is
obtained from the nonexponentiality of $g^{\left( 2\right) }\left( \tau
\right) $ related to the exponentiality of the average decay rate at a
single scattering angle, the apparent
hydrodynamic radius and its distribution obtained at a scattering angle of 0
are different from the hydrodynamic radius and its distribution. 

\begin{table}[tbp]
\begin{center}
\begin{tabular}{|c|c|c|c|c|c|}
\hline
$\sigma /\left\langle R_{s}\right\rangle $ & \multicolumn{5}{|c|}{Scattering
angle} \\ \hline
& 0 & 30$^{\text{o}}$ & 60$^{\text{o}}$ & 90$^{\text{o}}$ & 120$^{\text{o}}$
\\ \hline
0.06 & 55.98 & 55.97 & 55.94 & 55.91 & 55.87 \\ \hline
0.1 & 57.65 & 57.62 & 57.55 & 57.45 & 57.35 \\ \hline
0.2 & 64.62 & 64.51 & 64.22 & 63.81 & 63.40 \\ \hline
0.3 & 74.15 & 73.90 & 73.22 & 72.27 & 71.30 \\ \hline
0.4 & 85.03 & 84.56 & 83.28 & 81.48 & 79.67 \\ \hline
0.5 & 96.63 & 95.84 & 93.69 & 90.71 & 87.74 \\ \hline
\end{tabular}
\makeatletter
\par
\makeatother
\end{center}
\caption{Values of $R_{h,app}$ obtained at different scattering angles for
the simulated data produced based on the mean static radius 50 nm and
different standard deviations using the intensity-weighted average diffusion
coefficient.}
\label{averageD}
\end{table}

\section{CONCLUSION}

The nonexponentiality of $g^{\left( 2\right) }\left( \tau \right) $ is
determined by the particle size distribution and scattering angle. In
general, the effects of the particle size distribution are small on the
deviation between an exponentiality and $g^{\left( 2\right) }\left( \tau
\right) $ and very large on the initial slope of the logarithm of $g^{\left(
2\right) }\left( \tau \right) $ and the effects of the scattering angle are
determined by the particle size distribution and mean particle size. Under
some conditions, the nonexponentiality of $g^{\left( 2\right) }\left( \tau
\right) $ is greatly influenced by the scattering angle. The values of the
apparent hydrodynamic radius are a function of the particle size
distribution and scattering angle. The wider the particle size distribution,
the larger the value of the apparent hydrodynamic radius. The apparent
hydrodynamic radius and its distribution obtained using the cumulants method
are different from the hydrodynamic radius and its distribution.

Fig. 1. The differences between the lines and plots of $\ln \left( \left(
g^{\left( 2\right) }\left( \tau \right) -1\right) \right) $ as a function of
the delay time $\tau $ are explored at a temperature of 298.5 K and
scattering angles 30$^{\text{o}}$, 60$^{\text{o}}$, 90$^{\text{o}}$, 120$^{%
\text{o}}$ and 150$^{\text{o}}$. The symbols show the experimental data and
the lines show the linear fitting to the experimental data respectively. The
results for Latex-1 and Latex-2 are shown in a and b respectively.

Fig. 2. The differences between the lines and plots of $\ln \left( \left(
g^{\left( 2\right) }\left( \tau \right) -1\right) \right) $ as a function of
the delay time $\tau $ are investigated at a temperature of 312.6 K and
scattering angles 30$^{\text{o}}$, 60$^{\text{o}}$, 90$^{\text{o}}$, 120$^{%
\text{o}}$ and 150$^{\text{o}}$. The symbols show the experimental data and
the lines show the linear fitting to the experimental data respectively. The
results for PNIPAM-1 and PNIPAM-2 are shown in a and b respectively.

Fig. 3. The differences between the lines and plots of $\ln \left( \left(
g^{\left( 2\right) }\left( \tau \right) -1\right) \right) $ as a function of
the delay time $\tau $ are investigated at a temperature of 302.2 K and
scattering angles 30$^{\text{o}}$, 50$^{\text{o}}$ and 70$^{\text{o}}$. The
symbols show the experimental data and the lines show the linear fitting to
the experimental data respectively. The results for PNIPAM-1 and PNIPAM-2
are shown in a and b respectively.

Fig. 4. The fit results of $g^{\left( 2\right) }\left( \tau \right) $ for
the simulated data produced based on the mean static radius 50 nm and
standard deviations 10 nm at a scattering angle of 90$^{\text{o}}$. The
circles show the simulated data, the line represents the fit results
obtained using Eq. \ref{cum} and the diamonds show the residuals $\left(
y_{i}-y_{fit}\right) /\sigma _{i}$. The results for $\mu _{2}=0$ and $\mu
_{2}\neq 0$ are shown in a and b respectively.

Fig. 5. The differences between the lines and plots of $\ln \left( \left(
g^{\left( 2\right) }\left( \tau \right) -1\right) /\beta \right) $ as a
function of the delay time $\tau $. The symbols show the simulated data and
the lines show the linear fitting to the simulated data respectively. The
results for the simulated data at scattering angles 90$^{\text{o}}$ and 150$%
^{\text{o}}$ are shown in a and b respectively.

Fig. 6. The differences between the lines and plots of $\ln \left( \left(
g^{\left( 2\right) }\left( \tau \right) -1\right) /\beta \right) $ as a
function of the delay time $\tau $. The symbols show the simulated data and
the lines show the linear fitting to the simulated data respectively. The
results for the simulated data at scattering angles 30$^{\text{o}}$ and 90$^{%
\text{o}}$ are shown in a and b respectively.

Fig. 7. The differences between the lines and plots of $\ln \left( \left(
g^{\left( 2\right) }\left( \tau \right) -1\right) /\beta \right) $ produced
based on the mean static radius 50 nm with the standard deviations 3, 10 nm
and 25 nm and 120 nm with the standard deviations 8 nm, 24 nm and 60 nm at
the scattering angles 30$^{\text{o}}$, 90$^{\text{o}}$ and 150$^{\text{o}}$
as a function of the delay time $\tau $, respectively. The symbols show the
simulated data and the lines show the linear fitting to the simulated data
respectively. The results for the simulated data produced using the mean
static radii 50 nm and 120 nm are shown in a and b respectively.

Fig. 8. The differences between the lines and plots of $\ln \left( \left(
g^{\left( 2\right) }\left( \tau \right) -1\right) /\beta \right) $ produced
based on the mean static radius 120 nm and standard deviation 60 nm at the
scattering angles 30$^{\text{o}}$, 60$^{\text{o}}$, 90$^{\text{o}}$, 120$^{%
\text{o}}$ and 150$^{\text{o}}$ as a function of the delay time $\tau $,
respectively. The symbols show the simulated data. The results for the
simulated data at different scattering angles are shown in a and b
respectively.

Fig. 9. The differences between the lines and plots of $\ln \left( \left(
g^{\left( 2\right) }\left( \tau \right) -1\right) /\beta \right) $ produced
based on the mean static radius 120 nm and standard deviation 60 nm at the
scattering angles 30$^{\text{o}}$, 60$^{\text{o}}$, 90$^{\text{o}}$, 120$^{%
\text{o}}$ and 150$^{\text{o}}$ as a function of the delay time $\tau $
assuming $P\left( q,R_{s}\right) =1$, respectively. The symbols show the
simulated data and the lines show the linear fitting to the simulated data
respectively. The results for the simulated data at different scattering
angles are shown in a and b respectively.


\begin{thebibliography}{99}
\bibitem{re1} D. E. Koppel, J. Chem. Phys. \textbf{57}, 4814(1972).

\bibitem{re2} C. B. Bargeron, J. Chem. Phys. \textbf{61}, 2134(1974).

\bibitem{re3} J. C. Brown, P. N. Pusey and R. Dietz, J. Chem. Phys. \textbf{%
62}, 1136(1975).

\bibitem{re4} B. J. Berne and R. Pecora, \textit{Dynamic Light Scattering}
(Robert E. Krieger Publishing Company, Malabar, Florida, 1990).

\bibitem{re5} G. Bryant, S. Martin, A. Budi and W. van Megen, Langmuir 
\textbf{19}, 616(2003).

\bibitem{re6} P. N. Pusey and W. van Megen, J. Chem. Phys. \textbf{80},
3513(1984).

\bibitem{re7} Y. Sun arxiv.org/abs/physics/0511159.

\bibitem{re8} Y. Sun arxiv.org/abs/physics/0511160.

\bibitem{re9} Y. Sun arxiv.org/abs/physics/0511161.

\bibitem{re10} The ALV Manual of the Version for ALV-5000/E for Windows,
ALV-Gmbh, Germany, 1998.
\end{thebibliography}
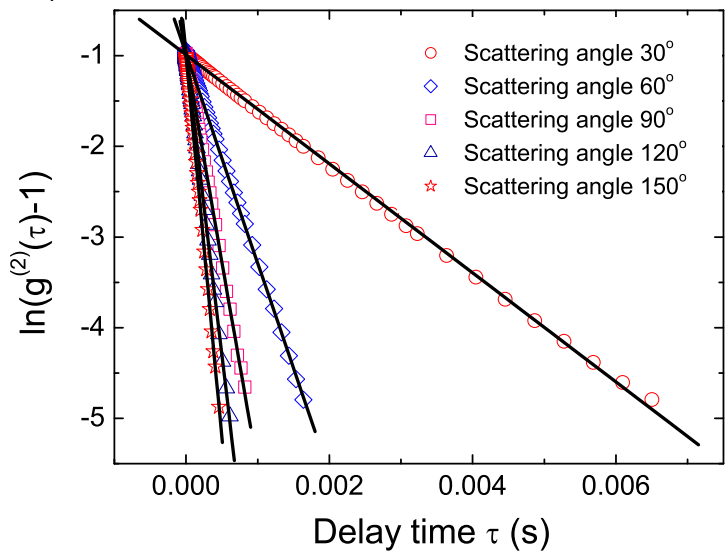
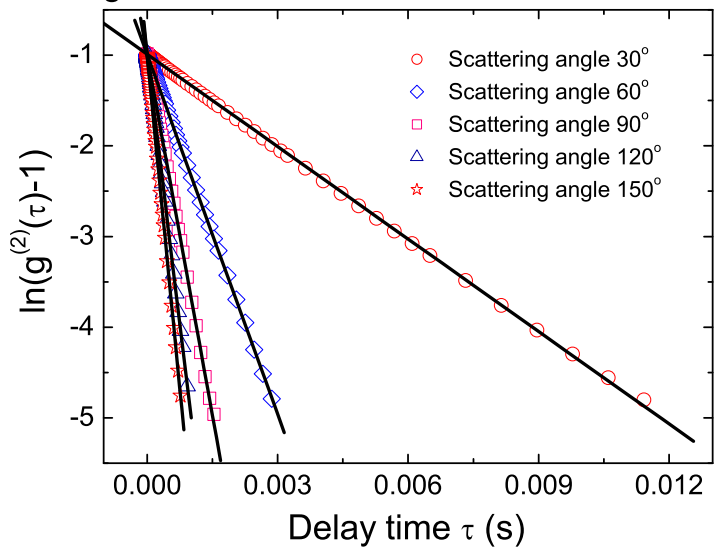
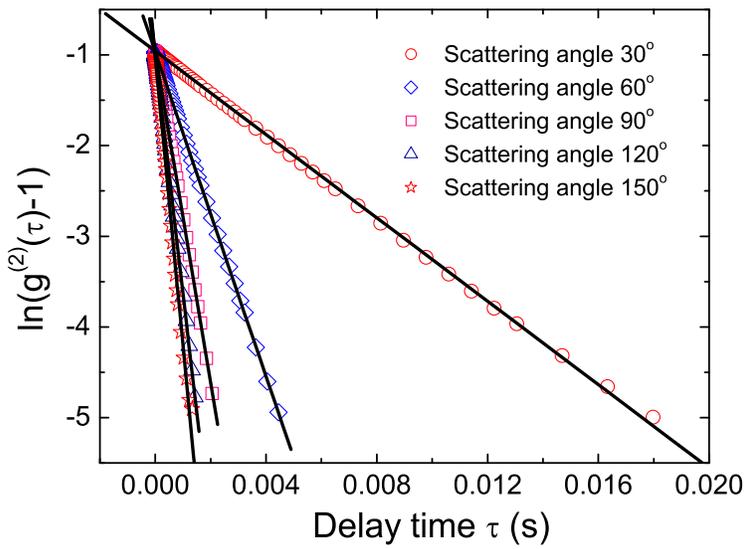
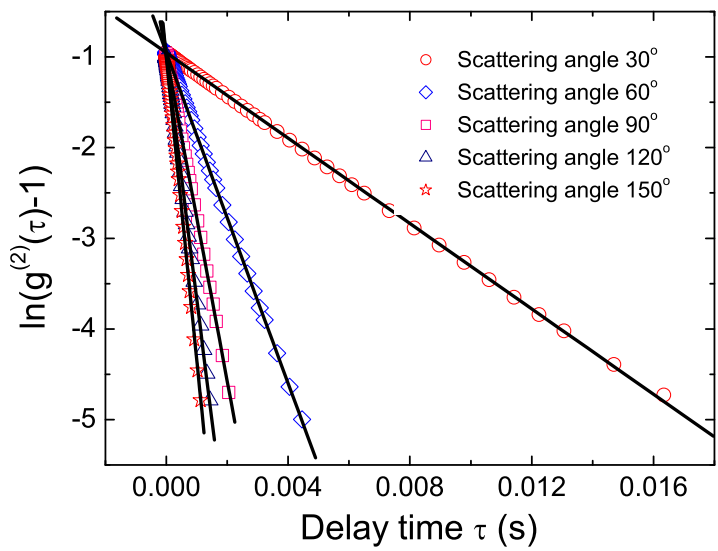
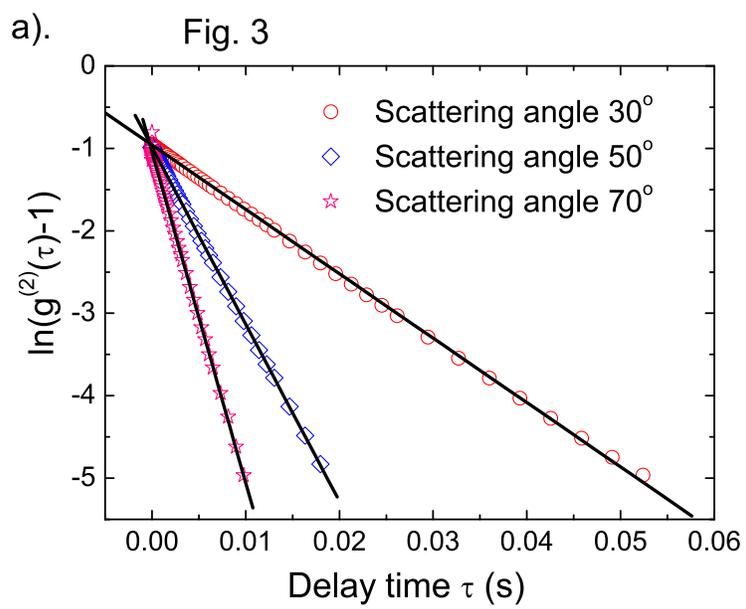
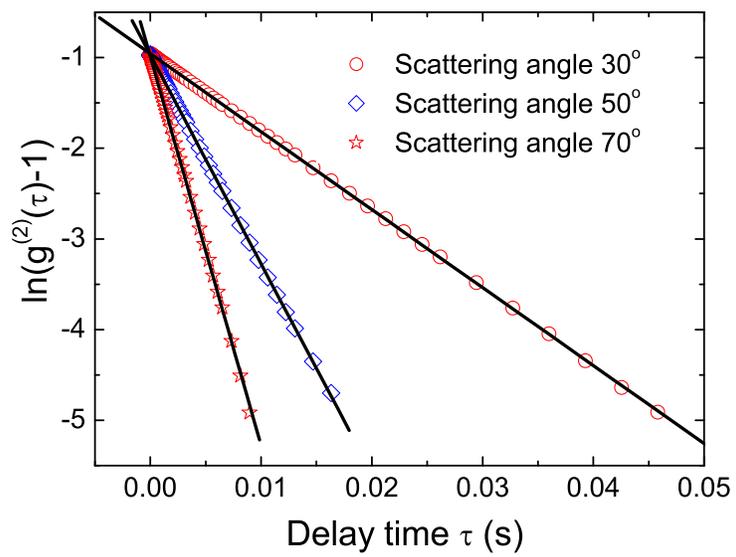
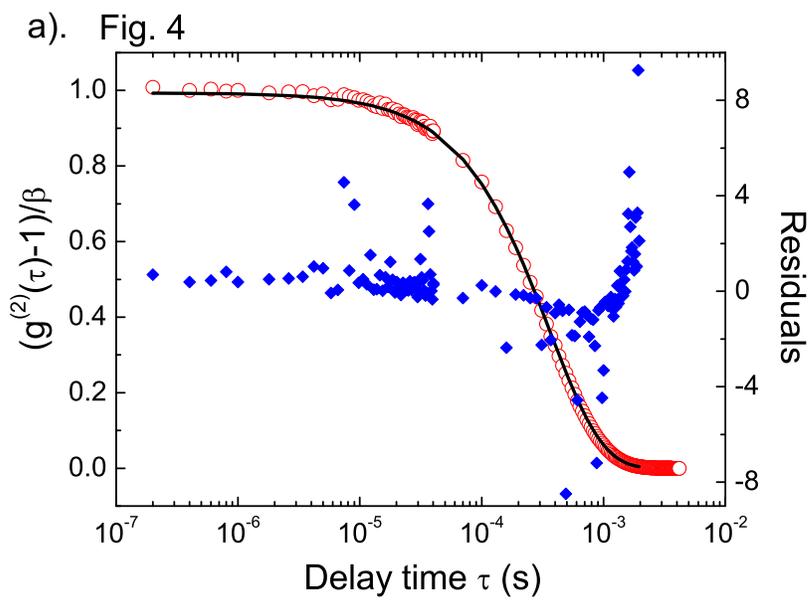
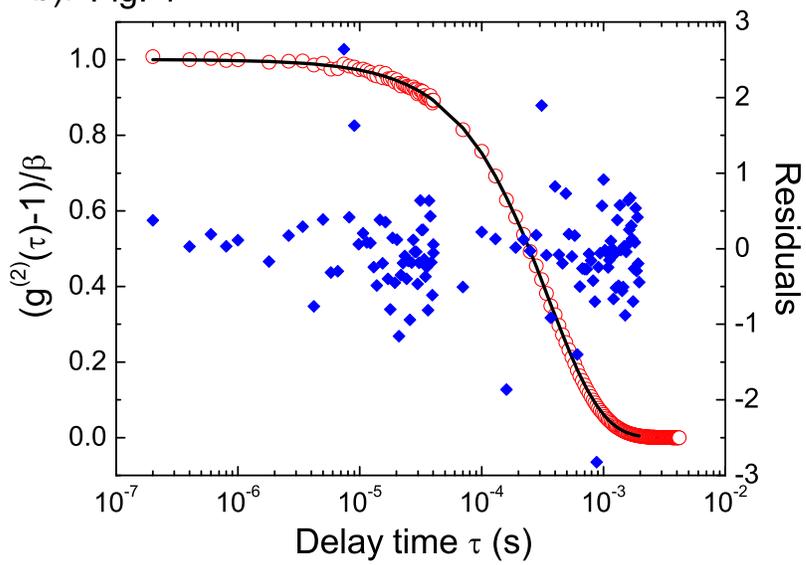
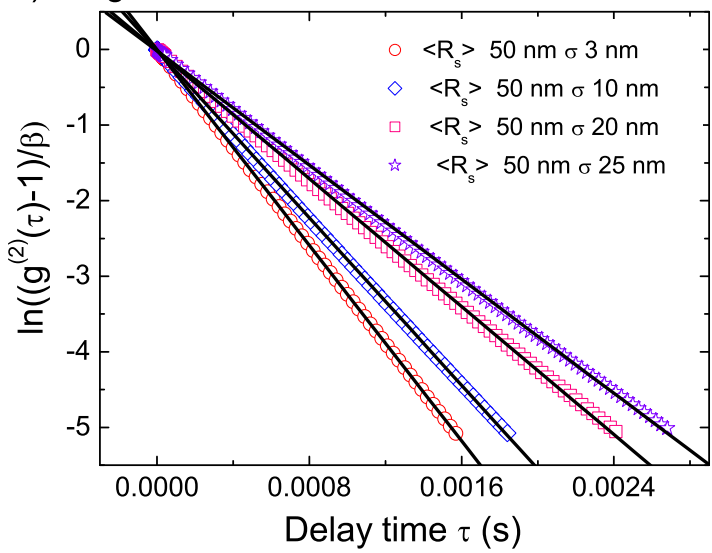
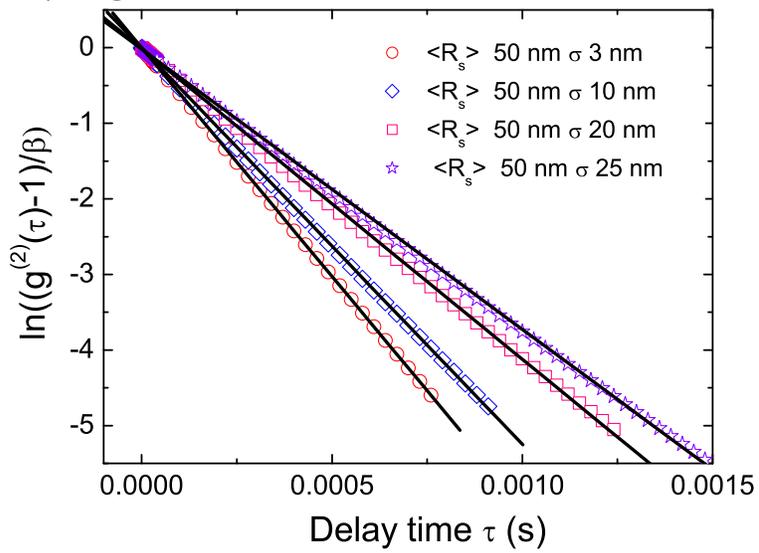
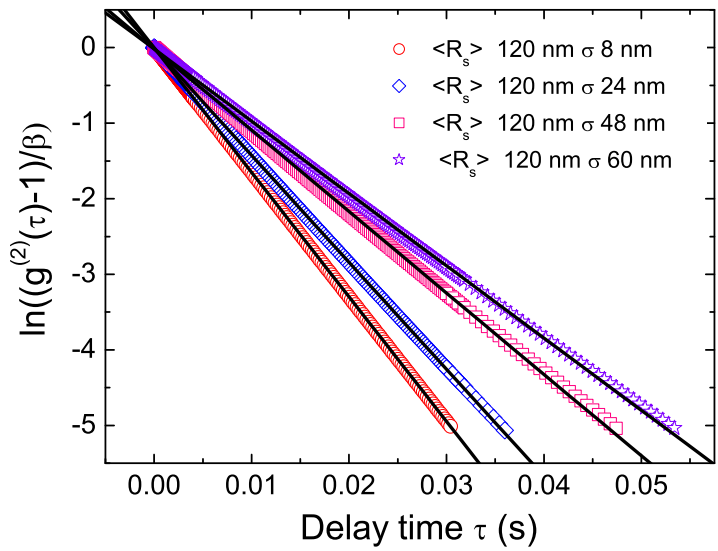
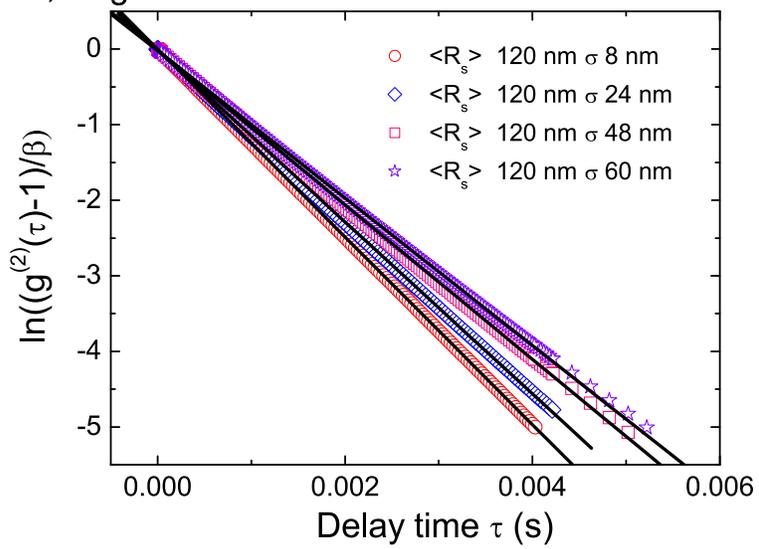
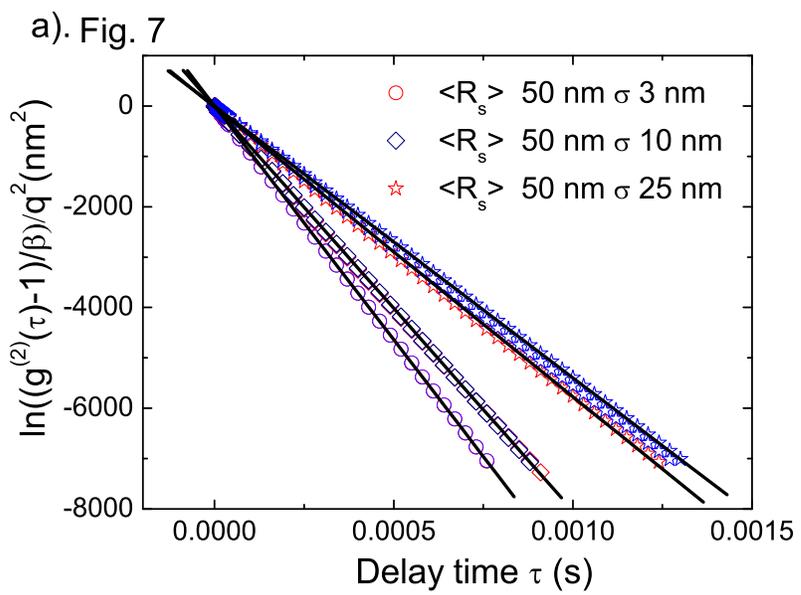
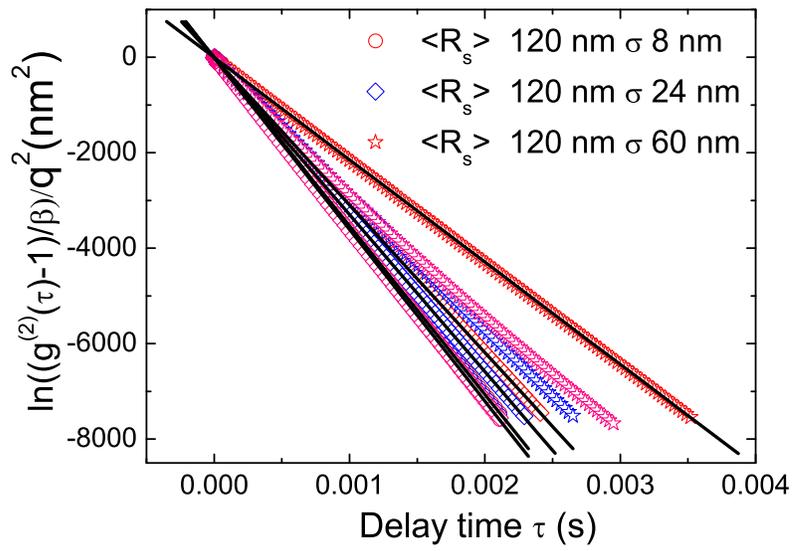
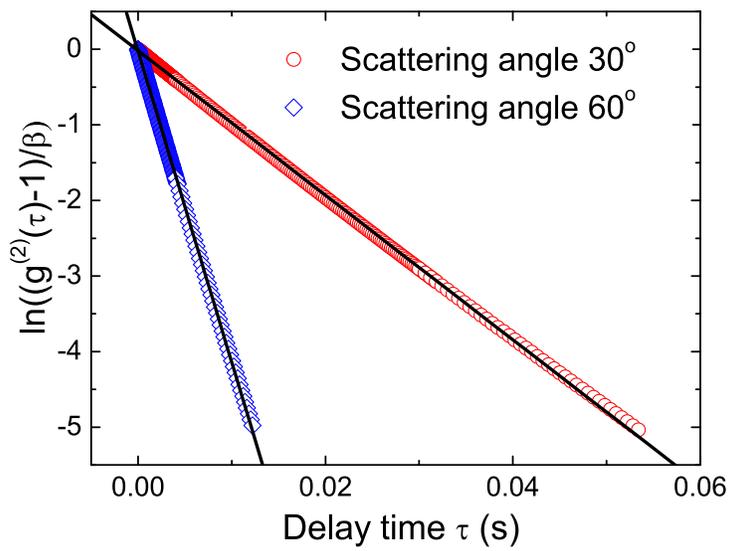
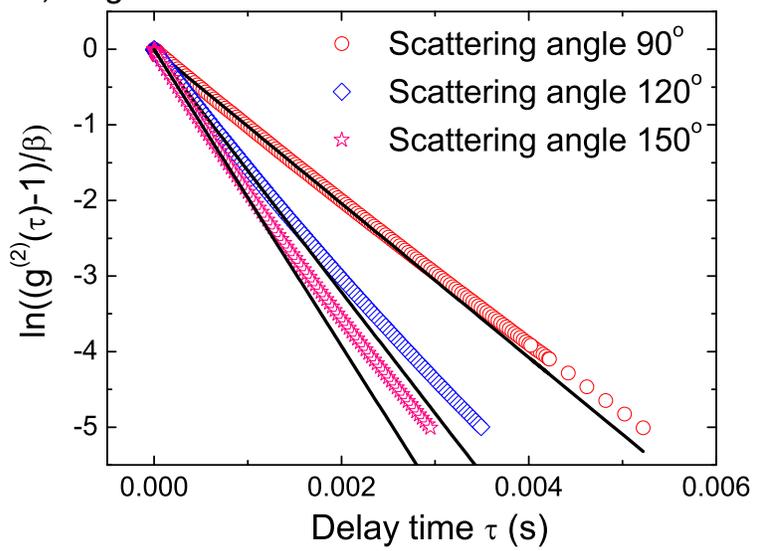
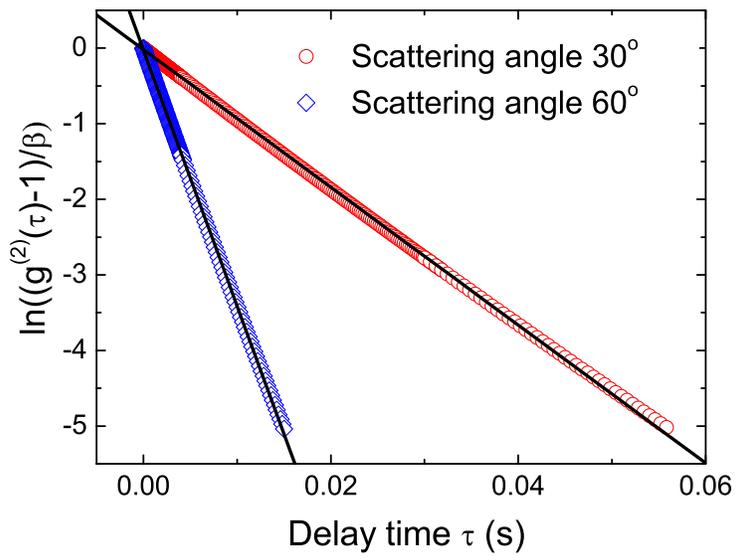
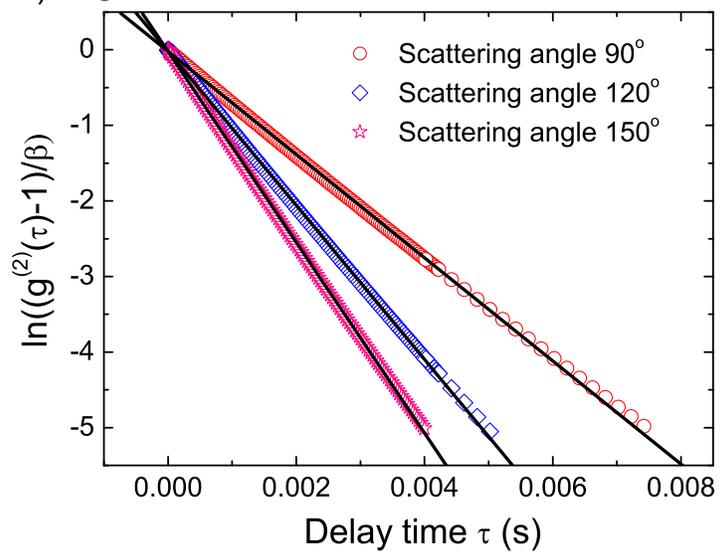
\end{document}